\documentclass[12pt,manuscript]{aastex}
\shorttitle{V1647 Ori in Outburst}
 
\begin{document}

\title{V1647 Ori (IRAS 05436$-$0007) in Outburst: the First Three Months}
\author{Frederick M.\ Walter}
\affil{Department of Physics and Astronomy, Stony Brook University,
Stony Brook NY 11794-3800\\fwalter@astro.sunysb.edu}
\author{Guy S.\ Stringfellow}
\affil{Center for Astrophysics and Space Astronomy,
        University of Colorado, Boulder CO 80309\\
        Guy.Stringfellow@Colorado.edu}
\author{William H.\ Sherry}
\affil{Department of Physics and Astronomy, Stony Brook University,
Stony Brook NY 11794-3800\\wsherry@astro.sunysb.edu}
\author{Angeliki Field Pollatou}
\affil{Department of Physics and Astronomy, Stony Brook University,
Stony Brook NY 11794-3800\\angeliki@astro.sunysb.edu}

\begin{abstract}
We report on photometric (BVRIJHK) and low dispersion spectroscopic
observations of V1647~Ori, the star that drives McNeil's Nebula,
between 10 February and 7 May 2004.
The star is photometrically variable atop a general decline
in brightness of about 0.3-0.4~magnitudes during these 87 days. The spectra
are featureless, aside from H$\alpha$ and the Ca~II infrared triplet in
emission, and a Na~I D absorption feature. The Ca~II triplet line ratios
are typical of young stellar objects. The H$\alpha$ equivalent width
may be modulated on a period of about 60 days. The post-outburst extinction
appears to be less than 7~mag.
The data are suggestive of an FU~Orionis-like event, but further monitoring
will be needed to definitively characterize the outburst.
\end{abstract}
 
 
\section{Introduction}

The appearance of a new star is a noteworthy event.
Since the discovery of a newly visible nebulosity in the L1630
dark cloud (McNeil 2004), the eruptive object that
illuminates McNeil's nebula has generated a flurry of interest 
(Reipurth \& Aspin 2004; \'Abrah\'am et al.\ 2004;
Brice\~no et al.\ 2004; Vacca, Cushing, \& Simon 2004;
Andrews, Rothberg, \& Simon 2004). 
Such outbursts of young stars are believed to arise from enhanced accretion
events from a circumstellar accretion disk.
Two classes of eruptive young stellar objects have been previously
identified (Herbig 1977): the FU~Orionis objects (FUors) and the 
EX~Lupi-like EXors.
McNeil's nebula itself is variable,
and visible presumably only while the illuminating object is in outburst.
We report here on three months of
optical spectroscopic and optical and near-IR photometric monitoring beginning
immediately following the report of the discovery. The long term variations
we see may help clarify the nature of the eruptive object.

The eruptive object, V1647~Ori, is positionally
coincident with the IR source IRAS 05436$-$0007, the sub-mm source
OriB55smm (Mitchell et al.\ 2001), the mm source LMZ~12 (Lis, Menten, \&
Zylka 1999), and the near-IR point source 2MASS J05461313$-$000648. 
Lis et al.\ (1999) proposed that the pre-outburst source
was an embedded Class~0 object,
with a bolometric luminosity of about 2.7~L$_\odot$.
\'Abrah\'am et al.\ (2004)
contrasted the quiescent state of the object with the
observed properties of both the FUor and EXor classes.
Based primarily on spectral energy distributions (SEDs) they concluded
the object resembled more the FU~Orionis objects, because the
flat-spectrum indicates the presence of a circumstellar envelope. 
They inferred a bolometric luminosity of 5.6~L$_\odot$; the larger luminosity
is attributable to a more complete SED than used by Lis et al., including
additional 2MASS and ISO data. Their resulting SED is that of a 
Class~I/II source with a fairly massive circumstellar envelope.
Andrews et al.\ (2004) concur, 
but by ascribing different dust properties they find the 
envelope to be less massive, and a bolometric luminosity of 3.4~L$_\odot$.
Nevertheless, the disk mass derived in both cases is significantly higher than
found for most Class~II disks. 
Either way, the luminosity suggests that V1647~Ori is a low mass object.

Bricen\~o et al.\ (2004) show that the current outburst began between
28~October and 15~November 2003. They detected the object on three occasions
in 1999 with I$_c$ between 18.4 and 20.3. Eisl\"offel \& Mundt (1997)
noted the object on a deep I-band image obtained in 1995. 
The 2MASS JHK detections occurred in October 1999 (Reipurth \& Aspin 2004).
The nebula itself appeared on an image taken in 1966 (Mallas \& Kreimer 1978),
which suggests that the star was in outburst at that time.

Published spectra of the object show emission in the hydrogen lines
(H$\alpha$, Pa$\beta$, Pa$\gamma$, Br$\alpha$, Pf$\gamma$). H$\alpha$
(Reipurth \& Aspin 2004) and the Paschen lines (Vacca et al.\ 2004) show
P~Cygni profiles indicative of a strong outflow, with velocities up to
600~km/s. While classical T Tauri stars often show evidence for outflows
in H$\alpha$, the velocities are generally smaller and the absorption is 
seen on a much broader emission line rather than against the continuum.
Also in emission are the 2.3-2.5$\mu$m CO bands and various metallic lines
in the K band (Vacca et al.\ 2004).

Brice\~no et al.\ (2004) showed that the spectrum in the direction of HH~22
resembled that of an early~B star in February 2004.
The Herbig-Haro object HH~22 (Herbig 1974; Eisl\"offel \& Mundt 1997) lies in
the direction of the outflow channel illuminated by V1647~Ori. The HH object
itself has an emission line spectrum; the continuum is likely reflected off
nearby dust.
If the HH~22 region is illuminated by V1647~Ori,
the continuum would be the reflection spectrum
of the star in outburst.
However, the observations by Eisl\"offel \& Mundt (1997) appear to 
indicate that HH~22 was ejected by an as yet unidentified source to its East, 
as a jet structure is evident in that direction. It may be premature to
conclude that V1647~Ori illuminates HH~22;
the spectrum of HH~22 could be a composite of sources contributing to its
illumination.
Nonetheless, for A$_I$ assumed to be 7.2~mag, the early B spectrum suggests
an outburst luminosity of about 220~L$_\odot$.
Vacca et al.\ (2004) estimated A$_V\sim$11~mag, based on the
optical depth of the 3.0$\mu$m ice absorption feature using a
relation between the 3$\mu$m optical depth and A$_V$ (Whittet et al.\ 1988).

Andrews et al.\ (2004) show that the post-outburst SED, like the pre-outburst
SED, is that of a flat spectrum source. 
They conclude that the post-outburst
luminosity lies between 34 and 90~L$_\odot$. This is less than Brice\~no's
luminosity estimate primarily because they used a smaller extinction.

What is clear is that V1647~Ori is an eruptive object that is illuminating
McNeil's nebula; what is not clear is the nature of the outburst and the 
underlying central source.
The two primary classes of erupting low mass pre-main sequence
stars, the FU~Orionis objects and the EXors,
are both driven by accretion. 
The FU~Ori objects appear to be the consequence of large-scale
instabilities in the inner disk which result in lengthy (decades to
centuries long) outbursts, during which time the accretion luminosity of the
inner disk dominates (Hartmann \& Kenyon 1996).
EXors (e.g., Herbig et al.\ 2001)
are classical T~Tauri stars that undergo large
outbursts ($>$3 magnitudes) on timescales of weeks to months.
Brice\~no et al.\ (2004) state that the reflection spectrum resembles
those of the
FU~Orionis object V1057~Cyg shortly after outburst.
However, Reipurth \& Aspin (2004) note
that the near-IR spectra bear some similarity with those of EXors in
outburst. Our catalog of
eruptive pre-main sequence stars is still very small, and we
do not really know what behavior is typical for these objects.
A distinction between these two types of objects may be found in the long-term
spectrophotometric variations, which have not been addressed to date. That is
the aim of this work.

\section{Data Reduction and Analysis}
All the data we report here were obtained with the SMARTS\footnote{
SMARTS, the Small and Medium Aperture Research Telescope Facility, is
a consortium of universities and research institutions that operate the
small telescopes at Cerro Tololo under contract with AURA.}
facilities at
Cerro Tololo. All data were obtained by the SMARTS service observers.
All dates below refer to the civil date at the start of the night.

We obtained images of the McNeil Nebula field on 25 nights between 10 February
and 7 May 2004 (Table~\ref{tbl-phlog}).  
With one exception, the images were obtained with the
1.3m ANDICAM imager, which obtains simultaneous optical and near-IR images. 
We have useful near-IR images on 23 nights. On one night clouds prevented 
detection of the comparison star, and we have no near-IR data simultaneous
with the 0.9m images. We obtained U band images on the first 3 nights, but 
V1647~Ori is not visible in these images.
Exposure times with ANDICAM are 300~sec (B), 109~sec (V, R, I), 
160~sec (J), 48~sec (H), and 24~sec (K). The J, H, and K images are
dithered, with a 30\arcsec\ throw to 6 or 7 positions.

The optical ANDICAM data are processed (overscan subtraction and
flat-fielding) prior to distribution.
Because we obtained only single images through each filter each night,
the noise is dominated, in some cases, by cosmic rays and hot pixels.
Aside from a 2x2 rebinning, the IR data are delivered raw.
We generate the flat field images from the
dome flats obtained about every other night. 
We generate a sky image by taking the median of the dithered images.
We subtract the sky image from the individual frames, then shift and add the
individual frames. 

Standard practice is to use the ANDICAM for differential photometry,
but the SMARTS project routinely obtains images of optical and near-IR
standard fields on photometric nights to obtain at least a
zero-point for the photometry. 

We obtained a set of BVRI images with the 0.9m and the 2k CCD imager on
19 February 2004. The night was photometric. We processed the data using
standard IRAF techniques.
The residuals to the photometric solution are $<$0.01~mag.
The absolute photometry is presented in
Table~\ref{tbl-phot}.

The spectra (Table~\ref{tbl-splog}) were obtained with the RC spectrograph
on the 1.5m telescope. This is a slit spectrograph, with a 300\arcsec\
long slit oriented E-W. We observed through a 110$\mu$m (1.5\arcsec) slit. 
We obtained three images at each epoch in order to filter cosmic rays.
Each set of images is
accompanied by a wavelength calibration exposure.
We have developed a pipeline, written in IDL, to process the data.
We subtract the overscan, and divide by the normalized flat field image.
We generate a median image from the three images. We extract the spectrum,
both by using an unweighted boxcar extraction and by fitting a Gaussian profile
at each wavelength. In the boxcar extraction, the extraction slit width is
determined by fitting a Gaussian to the spatial profile, and the
background is measured to either side of the source.
We observed a spectrophotometric standard, LTT 4364, each night in order to
convert the counts spectrum to a flux spectrum. 
Due to seeing-related slit losses, we do not obtain absolute fluxes, but rather
use this to recover the shape of the continuum.

We obtained the astrometric solution by fitting the stars in the USNO~A2 
catalog visible in the 0.9m images. V1647~Ori is at
05:46:13.166 -00:06:04.64 (J2000), with uncertainties of less than 0.5~arcsec. 
This is consistent with the coordinates of the IR counterpart, variously
known as 2MASS~05461313-0006048, LMZ~12, and IRAS 05436-0007.

\subsection{Differential Photometry}

The differential photometry is hampered by the small number of comparison
stars available within the 6.3\arcmin\
ANDICAM field of view. Of the three stars visible in
the optical channel, two are known T Tauri stars (LkH$\alpha$~301 
[Cohen \& Kuhi 1979], and star~V of Brice\~no et al.\ 2004).
We confirm that both are photometrically variable.
The other star (J2000 coordinates = 05:46:22.426 -00:03:38.39)
appears to be steady to within 0.05~mag,
based on comparisons with photometric standards on photometric nights.
The BVRI magnitudes of this star, from the photometrically-calibrated
0.9m~image, are listed in Table~\ref{tbl-phot}. 

Because of possible contamination by the variable nebula, it is important to
understand the near-star environment and the nebular contribution to
the background.
To minimize contamination by the nebula, we use a 1.85\arcsec\ (5 pixel) radius
aperture to extract the source counts from the optical images.
The background is the median within an
annulus with inner and outer radii of 10 and 20 pixels, respectively. Any
nebular emission within the annulus is rejected by taking the median,
since the nebula occupies a small fraction of the annular area. We verified
that we are not oversubtracting the background
by examining the data within the annulus. The adopted background
level matches the observations to the south and east of V1647~Ori; to the
northwest diffuse emission from McNeil's Nebula dominates.

We cannot be confident that we are not undersubtracting the sky, since there
may well be significant nebular emission within the 1.85\arcsec\ 
extraction radius. There is spatially-asymmetric emission within the
1.85 to 3.7\arcsec\ gap. This is a southward continuation of the
extended nebula in
towards V1647~Ori. To quantify this, we extracted cuts through the images on a
night with good seeing (23 March). These cuts passed through V1647~Ori with
a position angle of between 20$^\circ$ and 30$^\circ$. The surface brightness
of McNeil's Nebula peaks about 10\arcsec\ from V1647~Ori.
We extrapolated the surface brightness of the nebula linearly from about
5\arcsec\ north-northeast of the star
in order to estimate the degree of contamination
of the starlight by background nebulosity. 
Under this assumption we find that the maximum contribution of the
nebula within the inner 1.5\arcsec\ is
7\% of the total flux
in the I~band, 11\% at R, 23\% at V, and 13\% at B.
The relatively large contamination at V is due to the fading of the source; 
the relatively smaller contribution at B is due to the fading of the
nebulosity.

In the smaller (2.4\arcmin\ square) IR channel, there is
only one other star visible, at 05:46:11.68 -0:06:28.3 (J2000). This is
2MASS~05461162-0006279 ($J=13.94$, $H=12.19$, $K=11.20$), and by default it
is our comparison star.
We know nothing else about this star. We used a 2.5\arcsec\ (9 pixel) radius
extraction aperture for the photometry. The mean IR magnitudes of
our target are $J=10.9$, $H=8.9$, $K=7.3$, in good agreement with the 
near-IR magnitudes reported by Reipurth \& Aspin (2004).

The differential light curves are show in Figures~1 and~2.
In both the optical and near-IR there is a general downwards
trend. The trend is significant at 
$>$99.8\% confidence in all bands except $B$, where the star is faintest and
contamination from the nebula is greatest.
Over the 87 days we followed the star, the mean brightness decreased by
about 0.4~mag at $I$ and 0.3~mag at $K$.
We detect no significant color changes; over this time the mean
$V-K$ color reddened by 0.18$\pm$0.16~mag. 

The general downward trend is superposed on a variable source. Only about 
30\% of the RMS scatter is attributable to any long-term trends.
There are
dips in the optical light curve around 7 March and 12 April. The dips are
most pronounced at the shortest wavelengths, and are not visible in the
near-IR, suggesting that they may be due to variable dust obscuration
with a change in A$_V$ of 0.2-0.3~mag.

In Table~\ref{tbl-phlog} we record the measured FWHM of the stellar PSF in
the I~band.
This is determined from a Gaussian fit to two stars, LkH$\alpha$~301 and
star V, and is a measure of the seeing averaged over the 109 second
integrations.
We see no correlations between the relative brightnesses and the seeing,
which suggests that the brightness variations are intrinsic to V1647~Ori
(or an unresolved compact nebula).

We repeated the aperture photometry with a smaller 3 pixel (1.1\arcsec) radius
aperture. The shape of the light curve is identical, within the errors,
to that extracted with the larger aperture. This is further evidence 
that the source of the variations is not McNeil's Nebula, but is
V1647~Ori, or perhaps a very compact nebula within 1\arcsec\ of the star.
The principal effect of any nebular contamination will be to dilute the 
amplitude of any stellar variations, and to the affect the inferred colors at
short wavelengths.

\subsection{Spectroscopy}

We obtained spectra on 10 nights between 13 February and 17 April 2004
(Table~\ref{tbl-splog}).
Eight are first order spectra using grating 47 and the GG~495 order
sorting filter to obtain 3.1\AA\ resolution between 5650 and 6970\AA.
This resolution is sufficient to resolve the H$\alpha$ line.
Two spectra were obtained using grating 13 unfiltered, which yields a 17.2~\AA\
resolution spectrum over nearly the full optical spectral region from 3150 to
9370\AA.

The spectra show few features. H$\alpha$ is in emission
(Figure~3), with a P~Cygni
absorption feature, as reported by Reipurth \& Aspin (2004). Our resolving
power is about 40\% of theirs, so we are not sensitive to any structure in the
outflow profile. 

The equivalent width of the H$\alpha$ emission line varies from $-$30\AA\ to
$-$48\AA\ (see Table~\ref{tbl-ha}),
with formal measurement uncertainties of order 0.1\AA. Systematic 
uncertainties in the placement of the weak continuum dominate the
uncertainties. Many of the spectra were taken before the end of twilight, with
the sky brightness comparable to that of the star. We estimate that the
systematic errors in the equivalent width are about $\pm$1\AA.
The $-$32\AA\ equivalent width reported by
Reipurth \& Aspin (2004) on 14 February is identical to the
equivalent width we observed the previous night.
The equivalent width of the emission line varies at the $\pm$20\% level.
The time variation (Figure~\ref{fig-haew})
is suggestive of an oscillatory or periodic variation,
but we sample too short a time interval to permit us to draw any conclusions.

We note that there appears to be an anticorrelation between the equivalent
widths of the P~Cygni absorption feature and of the H$\alpha$ emission line.
A modulation of the wind velocity or optical depth could yield this sort
of anticorrelation. However, our spectral resolution is insufficient to
permit any meaningful line profile fitting.

We do not see any emission at H$\beta$. We can trace the continuum down to
about 4300\AA. The limiting equivalent width, based on the noise in the
4860\AA\ continuum, is about 5\AA.

The Na~I D lines
are seen in absorption, apparently saturated, with an equivalent width of 
14\AA. This is consistent with the large extinction seen towards the star.

The Ca~II IR triplet is in emission (Table~\ref{tbl-ca}, Figure~5).
The star shows the unique
$\approx$2:2:1 pattern seen only in T~Tauri stars and Herbig Ae/Be stars
(Hamann \& Persson 1992a, b). These line ratios are inconsistent with the
1:9:5 $gf$ values, and are inconsistent with either pure optical depth effects
or simple chromospheric models (Hamann \& Persson 1992a).
The presence of the Ca~IR triplet lines in this emission pattern confirm
that the star is a pre-main sequence star, but we cannot use the fluxes to
determine whether the underlying star is a low mass classical T~Tauri star
or a more massive Herbig~Ae star.

\section{Discussion}

The discovery of a new star was once considered a portent, often with
dire consequences. In today's more rational cosmology,
a new star offers an opportunity to probe some of the earliest and
best hidden processes of the building of stars. It has been suggested that
rapid disk accretion via the FU~Orionis episodes is a major source of the
mass that builds low mass stars, since the inferred mass accreted in these
events is comparable to that accreted at a lower rate over the 10$^6$ year
pre-main sequence phase of a classical T~Tauri star
(see review by Hartmann \& Kenyon 1996).
As our characterizations of these episodes, and our inferences concerning
outburst timescales and recurrence rates, are based on a small and
inhomogeneous sample (5--9 stars as of the Hartmann \& Kenyon review),
every new outbursting pre-main sequence star can contribute a great deal to our
understanding of these phenomena.

The pre-outburst star had a luminosity between 2.7 and 5.6~L$_\odot$,
which is typical of a low mass classical T~Tauri star
(Lis et al.\ 1999, \'Abrah\'am et al.\ 2004, and Andrews et al.\ 2004).
This is based on integrating the 1$\mu$m - 1.3mm spectrum.
The small discrepancy in the luminosity is the consequence of
the different choices of A$_V$, and inclusion of slightly different
data sets.

%The overall properties of V1647~Ori post-outburst
%are generally consistent with classification as an FU~Orionis object.
If V1647~Ori is an FU~Orionis object, then the
reflection spectrum from the vicinity of HH~22 would mirror
the inner edge of the accretion disk. An A-type spectrum with
T$_{eff}\sim$8000K,
as seen in V1057~Cyg shortly after outburst, is consistent with this
scenario. This temperature, and the $\sim$~50~L$_\odot$ luminosity
(Andrews et al.\ 2004), suggests an inner disk radius
R$_i$=3.7 [(L/50~L$_\odot$)/(T/8000~K)$^2$]$^{-\frac{1}{2}}$~R$_\odot$,
and a mass accretion rate 
\.M~=~3$\times$10$^{-5}\frac{\rm L}{50~{\rm L}_\odot}\frac{{\rm R}_i}{3.7~{\rm R}_\odot}\frac{0.5~{\rm M}_\odot}{\rm M}~$M$_\odot$/yr.
For the range of luminosities and temperatures being
considered, these mass accretion rates and inner disk radii are 
consistent with, though typically lower than,
those of known FUors (Hartmann \& Kenyon 1996).
%Since several unknown intrinsic properties (dust properties, reddening, 
%complete contemporaneous coverage of the SED, etc.) feed into the determination
%of the pre- and post-outburst luminosity, we note that a lower outburst
%luminosity renders classification as an FUor more problematic.

\subsection{Limits on the Spectral Type of V1647~Ori}

With a luminosity of only
a few L$_\odot$ before the outburst, V1647~Ori was likely
of spectral type K--M, although the optical and near-IR spectrum may have
been heavily veiled by quiescent accretion.
Brice\~no et al.\ (2004) make the case that the reflection spectrum in
the vicinity of HH~22,
which resembles that of an 
early B star or A star, is the spectrum of V1647~Ori in outburst,
seen via a clear channel through the surrounding dust. The hot blue photosphere
could be that of a Herbig~Ae star (unlikely, given the pre-outburst luminosity)
or the luminous, optically thick inner part of the accretion disk.
We have examined our spectra in hopes of obtaining some clarification.

As mentioned earlier, the spectra are featureless, with the exception
of H$\alpha$ and the Ca~IR triplet in absorption, and the Na~I~D lines in
absorption.
We see no evidence of TiO absorption bands, which means that, in the
absence of overwhelming veiling, the spectral type of the photosphere can
be no later than K7. The limiting equivalent width of narrow absorption lines
in our coadded spectrum (Figure~\ref{fig-ha}) is 0.7\AA. We would not expect to
detect narrow lines from any star hotter than about spectral type K0, since 
the lines in the red are weak in the F-G stars. In the presence of significant
veiling, we can say even less.

If the spectrum is that of an A-F supergiant (the low density photosphere of
the hot disk), as was V1057~Cyg, the O~I $\lambda$7774\AA\
triplet should be prominent in absorption. There is a possible absorption
line here (W$_\lambda$=2.6\AA ), but it is near the noise level in the coadded
low resolution spectum and, absent
further information, is not significant.

The observed B through K colors of the star in outburst are more consistent
with a heavily reddened hot star than with a cool star. 
We unreddened the observed colors of V1647~Ori
using the parameterizations of Cardelli, Clayton, \& Mathis (1989). 
R, the ratio of total to selective extinction, was a selectable parameter.
We then compared the
unreddened colors to template stars AB~Aur, a Herbig Ae star, and T~Tau, 
the prototypical low mass pre-main sequence star.
The observed B through K colors are somewhat
consistent with the observed colors of the Herbig Ae star AB~Aur,
for R$\sim$5 and A$_V\sim$7$-$8~mag. A star with the colors of T Tauri requires
a much smaller A$_V$ of $\sim$4~mag.
This is not to say that V1647~Ori is a hot star; rather it is consistent with
the post-outburst light being dominated by a hot photosphere, perhaps an
inner accretion disk, as in the case of the FU~Orionis objects.
%This is consistent with Brice\~no's scenario.
%Alternatively, if the reflection spectrum is due to another 
%illuminating source, then a low-mass pre-main-sequence stellar spectrum would
%remain viable, indicating a significantly lower reddening during outburst.

%R    Th    Av   Tc     R    Th    Av   Tc
%5    4000  2.5  1300   3.1  4000  1.7  1250
%5    8000  6.7  1350   3.1  8000  4.5  1400
%5   20000  8.7  1450   3.1 20000  6    1500
%5   40000  9.2  1450   3.1 40000  6.4  1500

\subsection{The Extinction Towards V1647~Ori}

Estimates of the extinction towards V1647~Ori are highly uncertain, because
the shape of the underlying spectrum is unknown, and because there is no
constraint on R, the ratio of total to selective extinction.
While we have no $a~priori$ information on the value of R, we favor a large
value of R. R$\sim$5, indicative of large grain sizes, is commonly found
for embedded objects, and in regions of star formation (e.g., 
Savage \& Mathis 1979; Cardelli \& Clayton 1988), where large grains may
contribute to significant gray extinction. 

\'Abrah\'am et al.\ (2004) dereddened the pre-outburst
star on the $J-H,H-K$ color-color
diagram, apparently using R=3.1, until it lay on the unreddened
classical T~Tauri star locus. They concluded that A$_V$=13~mag. 
Had they dereddened the star using R=5.5, they would have estimated
A$_V$ to be about 2 magnitudes less. 

Vacca et al.\ (2004) used the empirical relation between $\tau$(3.1$\mu$m) and
A$_V$ (Whittet et al.\ 1988) to estimate A$_V$=11~mag. There is significant
scatter around this relation. In particular, the deeply embedded source
HL~Tau has $\tau$(3.1$\mu$m) = 0.85 (larger than V1647~Ori) but A$_V$ of
only 6-8 mag. The Whittet et al.\ (1988) relation was calibrated for stars
in the Taurus clouds, and there is evidence that the $\tau-$A$_V$
relations (and by inference the grain properties) vary between different
star formation regions (Teixeira \& Emerson 1999).

Brice\~no et al.\ (2004) state that the optical continuum slope is consistent
with A$_V$=8--10~mag, for an underlying A--B star photosphere.

We use the observed SED in the optical and near-IR to constrain A$_V$,
on the assumption that emission in the optical is thermal, and consists of a
single dominant temperature component. For a hot blackbody,
we find that A$_V<$9.7~mag for R=5 and A$_V<$6.8~mag for R=3.1.
In the more realistic case where the temperature is 8000--10,000K
(consistent with
a hot inner accretion disk or an A~star photosphere), we can place upper
bounds on A$_V$ of 6.7 or 4.5~mag for R=5 and 3.1, respectively.

Reipurth \& Aspin (2004) find that the extinction apparently decreased by
4.5 mag between the pre- and post-outburst apparitions.
Following the \'Abrah\'am et al.\ (2004) approach to de-reddening,
the difference between our outburst J-H color and the pre-outburst 2MASS 
color yields $\Delta$(J-H)=3.6 mag, or a decrease of A$_V \sim$5.4~mag during
outburst.
The possibility of reprocessing by scattering to longer
wavelengths makes this a minimum estimate. These two determinations
are consistent in indicating that the extinction decreased significantly
during outburst, with A$_V$ decreasing from about 11 to about 6~mag.
Either a great deal
of dust was sublimated during the outburst, or the wind blew a large hole in
the circumstellar envelope.

\subsection{The SED in the Optical and Near-IR }

We can fit the optical and near-IR (B through K)
spectral energy distribution as the sum of
two blackbody components. Figure~6 shows the case where the hotter
blackbody is 8000~K and R is 5.0.
A$_V$ and the temperature of the cooler blackbody are fit by eye. Given these
assumptions, the fit is unique, but we are able to obtain good fits
for any temperature exceeding 4000~K. The temperature
of the cool component always lies between 1200K and 1500K. The inferred
extinction, which makes the hot blackbody match the BVRI data, varies from
1.7~mag (R=3.1, T$_{BB}$=4000K) to 9.7 (R=5.0, T$_{BB}$=100,000K).
Physically these two components represent the boundary layer or inner
accreting material, and the inner part of the dust
disk, but don't address cooler material further from the star.

These two blackbody fits are consistent with the results of matching
the spectral energy distributions of AB~Aur and T~Tau, reported above.
The two-blackbody fits lie below the L$^\prime$ and M$^\prime$
fluxes reported by Vacca et al.\ (2004), as expected for a flat spectrum
source.

We can also fit the same data with a hot blackbody and a disk
model (with temperature decreasing as r$^{-\frac{3}{4}}$). 
The extinction is set by matching the hot blackbody to the optical points,
and is identical to that of the two-blackbody case.
Again, the model is
not unique, but we obtain a good fit with an inner dust disk temperature of
1450K and an inner disk radius of 12 R$_\odot$. This disk model
reproduces the L$^\prime$ and M$^\prime$ fluxes.
These values for the disk temperature and inner radius
are reasonable for a pre-main sequence circumstellar disk.

\subsection{FUor or EXor?}

Brice\~no et al.\ (2004) note that early spectra of the FU~Orionis object
V1057~Cyg
have H$\alpha$ profiles similar to those seen in V1647~Ori, but
Reipurth \& Aspin (2004) argue that the spectra are unlike those of mature
FUors. They suggest that V1647~Ori may be an example of an EXor.
While EXors and
FUors have comparable outburst amplitudes, the lengths of the
outbursts are very different, with FUors staying bright for decades. 
Over the course of our observations, V1647~Ori has faded by about 0.3--0.4~mag.

V1647~Ori was apparently bright around 1966. This recurrent nature, and the
relative rapidity of the fading (about 3-4 times as fast as V1057~Cyg --
see Geiseking 1973), are unlike any known FUor but is more
typical of the EXors.
V1647~Ori appears to have varied less erratically than 
did EX~Lup during the 87 days following any of its peaks during the 1993-1994
outburst (Herbig et al.\ 2001).
The 1955-1956 outburst of EX~Lup
took roughly 3 months to rise to maximum light, varied significantly 
about maximum light for about 3 months, then rapidly declined to its 
pre-outburst state within a matter of weeks (e.g., Herbig 1977). 
While the rise time and duration
at maximum light mimics that seen for V1647 Ori, the significant
variations seen in the EXor at maximum are not replicated.

We thus do not know how to categorize the star,
other than to say that it appears to
be a young accreting low mass pre-main sequence star in outburst.
V1647~Ori has brightened, and at the same time the extinction has decreased.
This may be telling us that there is a continuum of behaviors of
accreting low mass pre-main sequence stars, many of which we have yet to
see.

\section{Conclusions}

Based on the optical and near-IR photometry, and the optical spectra,
the extinction to V1647~Ori appears to be much smaller than
claimed elsewhere,
with a post outburst A$_V$ about 6.5~mag.
We suggest that R is close to 5. To reconcile $\tau$(3$\mu$m) with
A$_V$, we suggest that the dust grains are
more like those near HL~Tau than the ``typical'' T~Tauri star in Taurus.

After only three months of monitoring, it is not clear whether V1647~Ori
is best categorized as
a FUor or an EXor, or something in-between.
We know few members of either class, there is much variance among members of
each class, 
and our view of how they should behave may be severely biased.
However, the preponderance of the evidence to date
suggests that this outburst is
more like an FU~Orionis outburst.

Continued monitoring of V1647~Ori and McNeil's Nebula
when it reappears from behind the Sun late this summer
may ultimately tell us how best to characterize this outburst. 
However we finally decide to categorize this event,
this new star does portend a better understanding of the
nature of accretion-induced outbursts in pre-main sequence stars.

\acknowledgments

We are grateful for the support of Dean of Arts and Sciences J.~Staros,
Provost R.~McGrath, and Vice President for Research G.~Habich, all of
Stony Brook University, for providing partial support that enabled
Stony Brook's participation
in the SMARTS consortium. We thank the SMARTS service observers, J. Espinoza,
D. Gonzalez, A. Miranda, and A. Pasten, for taking the data,
and for their dedication to the SMARTS effort.
We thank C. Bailyn, the driving force behind the
SMARTS consortium, and R. Winnick, who accommodated our many requests to revise
the photometric timelines. We thank Tracy Beck for sharing her
insights into the extinction properties of highly embedded objects.
Finally we thank Bo Reipurth, the referee, for his very insightful comments.

This research was funded in part by NSF grant AST-0307454
to Stony Brook University.

\clearpage

\figcaption{The optical light curves in the B, V, R$_C$, and I$_C$ bands.
The differential magnitudes are relative
to a single comparison star (see text),
which appears constant to within 0.05 mag.
Error bars are $\pm$1~$\sigma$, based on counting statistics. The uncertainty
will be underestimated should the comparison star turn out to be variable.
The S/N is low on JD 2453065 because the field was observed
through clouds; the high point in the $V$ band should be discounted.
Uncertainties in the B band are large because the star is faint and because
there may be some contamination from McNeil's nebula. 
The dotted lines represent the mean magnitudes.
The downward trend in the
$V$, $R$, and $I$ bands is significant at $\alpha<$0.002.
}\label{fig-lcopt}  

\figcaption{The near-IR light curves.
The magnitudes are relative to comparison star
2MASS~05461162-0006279 (see text).
The dotted lines represent the mean magnitudes.
The downward trend in the
$J$, $H$, and $K$ bands is significant at $\alpha<$0.001.
}\label{fig-lcir}  

\figcaption{The mean H$\alpha$ profile, at 3.1\AA\ resolution.
This is the normalized sum of 7 spectra obtained between 13~February and
12~April. 
}\label{fig-ha}

\figcaption{The time variation of the H$\alpha$ equivalent width. Units are
\AA ngstroms. The formal
uncertainties on the equivalent width measurements are about 0.1\AA, but
systematic errors in the placement of the continuum are about an order
of magnitude larger. While the data are suggestive of an oscillatory or
periodic variation, only about 1.5 periods are sampled, and we offer no
interpretation of this.
}\label{fig-haew}

\figcaption{The Ca IR triplet, on 17~April (solid) and 21~March (dashed). The
continuua have been scaled to match. The equivalent widths changed by about
35\%, but the $\approx$2:2:1 pattern remains steady.
}\label{fig-ca}  

\figcaption{A simple two-blackbody fit to the optical (BVRI) and near-IR 
(JHK) magnitudes in February~2004.
We have fixed the temperature of the hotter blackbody at
8000~K, appropriate for an early A star photosphere. We set R, the ratio of
total to selective extinction, to 5.
A$_V$ and the temperature of the cooler blackbody are free parameters, fit by
eye. 
The lower solid curve is the 8000K blackbody; the dashed curve is the 1450K
blackbody. Both are reddened by A$_V$=6.7~mag. The upper solid curve is the
sum of the two reddened blackbodies.
}\label{fig-twobb}  

\clearpage
\begin{deluxetable}{lllll}
\tablecolumns{5}
\tablewidth{0pt}
\tablecaption{Photometric Log\label{tbl-phlog}}
\tablehead{
\colhead{Date (UT)} & \colhead{JD} & \colhead{FWHM\tablenotemark{a}} & \colhead{conditions} & \colhead{notes}}
\startdata
2004 Feb 11 & 2453046.604 & 1.7 & cirrus      & U image obtained\\ 
2004 Feb 12 & 2453047.674 & 1.8  & photometric & U image obtained\\
2004 Feb 13 & 2453048.651 & 1.2 & photometric & U image obtained\\      
2004 Feb 14 & 2453049.685 & 1.4 & photometric\\       %300.00/109
2004 Feb 20 & 2453055.590 & 1.2 & clouds \\
2004 Feb 21 & 2453056.533 & 2.1 & photometric & 0.9m images, no IR \\
2004 Feb 23 & 2453058.619 & 1.6 & patchy clouds \\
2004 Feb 26 & 2453061.564 & 2.0 & photometric\\
2004 Mar 01 & 2453065.573 & 1.4 & clouds\\
2004 Mar 02 & 2453066.604 & 1.9 & photometric\\
2004 Mar 05 & 2453069.580 & 1.2 & photometric\\
2004 Mar 08 & 2453072.524 & 1.3 & photometric\\
2004 Mar 12 & 2453076.543 & 1.6 & photometric\\
2004 Mar 17 & 2453081.580 & 1.3 & photometric\\
2004 Mar 23 & 2453088.492 & 1.0 & photometric\\
2004 Mar 30 & 2453094.502 & 1.9 & poor seeing\\
2004 Apr 02 & 2453097.513 & 1.6 & cirrus\\
2004 Apr 04 & 2453099.507 & 1.4 & patchy clouds & IR images unuseable\\
2004 Apr 06 & 2453102.499 & 1.1 & photometric\\
2004 Apr 07 & 2453103.489 & 1.3 & photometric\\
2004 Apr 09 & 2453105.504 & 1.4 & patchy clouds, cirrus\\
2004 Apr 10 & 2453106.487 & 2.2 & photometric\\
2004 Apr 12 & 2453108.491 & 1.9 & photometric\\
2004 Apr 17 & 2453113.495 & 1.3 & photometric\\
2004 May 07 & 2453133.447 & 1.5 & photometric\\
\enddata
\tablenotetext{a}{Measured FWHM of LkH$\alpha$~301 and star V, in arcsec. This
is a measure of the seeing. The 0.9m images on Feb 21 were not in
good focus.} 
\end{deluxetable}

\clearpage
\begin{deluxetable}{lrrrr}
\tablecolumns{5}
\tablewidth{0pt}
\tablecaption{Calibrated Photometry on 21 February 2004\label{tbl-phot}}
\tablehead{
\colhead{Star} & \colhead{B} & \colhead{V} & \colhead{R} & \colhead{I}\\
       & \colhead{$\pm$} & \colhead{$\pm$} & \colhead{$\pm$} & \colhead{$\pm$}}
\startdata
IRAS 05436-0007 & 18.421 & 16.931 & 15.704  & 14.209 \\
                &  0.085 &  0.023 &  0.009  &  0.006 \\ 
Comparison Star & 16.940 & 15.648 & 14.900  & 14.252 \\ 
                &  0.016 &  0.006 &  0.004  &  0.005 \\ 
\enddata
\end{deluxetable}

\clearpage
\begin{deluxetable}{llrrr}
\tablecolumns{5}
\tablewidth{0pt}
\tablecaption{Spectroscopic Log\label{tbl-splog}}
\tablehead{
\colhead{Date (UT)} & \colhead{UT}& \colhead{JD} & \colhead{dispersion} & \colhead{exposure}\\
                    &             &              & \colhead{(\AA /pix)}   & \colhead{(sec)}}
\startdata
2004 Feb 13 & 02:27 & 2453048.602 & 1.1 & 1800 \\
2004 Feb 23 & 01:14 & 2453058.551 & 1.1 & 1800 \\
2004 Mar 02 & 01:31 & 2453066.564 & 1.1 & 3300 \\
2004 Mar 05 & 01:08 & 2453069.548 & 1.1 & 3600 \\
2004 Mar 08 & 00:56 & 2453072.540 & 1.1 & 3600 \\
2004 Mar 21 & 00:01 & 2453085.500 & 5.7 & 3600 \\
2004 Mar 21 & 23:59 & 2453086.500 & 1.1 & 3900 \\
2004 Apr 06 & 00:30 & 2453101.521 & 1.1 & 3900 \\
2004 Apr 12 & 23:28 & 2453108.478 & 1.1 & 3600 \\
2004 Apr 17 & 23:20 & 2453113.473 & 5.7 & 3300 \\
\enddata
\end{deluxetable}

\clearpage
\begin{deluxetable}{lrr}
\tablecolumns{3}
\tablewidth{0pt}
\tablecaption{H$\alpha$ Measurements\label{tbl-ha}}
\tablehead{
\colhead{Date (UT)} & \multicolumn{2}{c}{W$_\lambda$(H$\alpha$)} \\
                    & \colhead{emission} & \colhead{absorption}}
\startdata
2004 Feb 13 & -32 & 3.3\\
2004 Feb 23 & \tablenotemark{a} & \nodata \\
2004 Mar 02 & -40 & 1.8\\ % +/- 1 A
2004 Mar 05 & -46 & 1.9\\
2004 Mar 08 & -48 & $<$0.6\\
2004 Mar 21.0 & -36 & 2.0\\ 
2004 Mar 21.9 & -37 & 1.5\\ 
2004 Apr 06  & -30 & 3.5\\ 
2004 Apr 12  & -37 & 1.5 \\ 
2004 Apr 17  & -45 & ---\\ 
mean         & -40 & 1.6\\
\enddata

\tablenotetext{a}{H$\alpha$ emission is clearly visible,
although no continuum was detected.}
\end{deluxetable}

\clearpage
\begin{deluxetable}{lrrr}
\tablecolumns{4}
\tablewidth{0pt}
\tablecaption{Ca IR Triplet Measurments\label{tbl-ca}}
\tablehead{
\colhead{Date (UT)} & \colhead{W$_\lambda$(8498)} 
                    & \colhead{W$_\lambda$(8542)} 
                    & \colhead{W$_\lambda$(8662)}}
\startdata
2004 Mar 21.0 &  -9.0 & -7.9 & -4.6\\ 
2004 Apr 17   & -11.1 & -10.4 & -7.8 \\ 
mean          & -10.2 & -9.4 &-6.0\\
\enddata
\end{deluxetable}

%%%%%%%%%%%%%%%%%%%%%%%%%%%%%%%%%%%%%%%%%%%%%%%%%%%%%%%%%%%%%%%%
%%%                  FIGURES
%%%%%%%%%%%%%%%%%%%%%%%%%%%%%%%%%%%%%%%%%%%%%%%%%%%%%%%%%%%%%%%%
 
\begin{figure}
\plotone{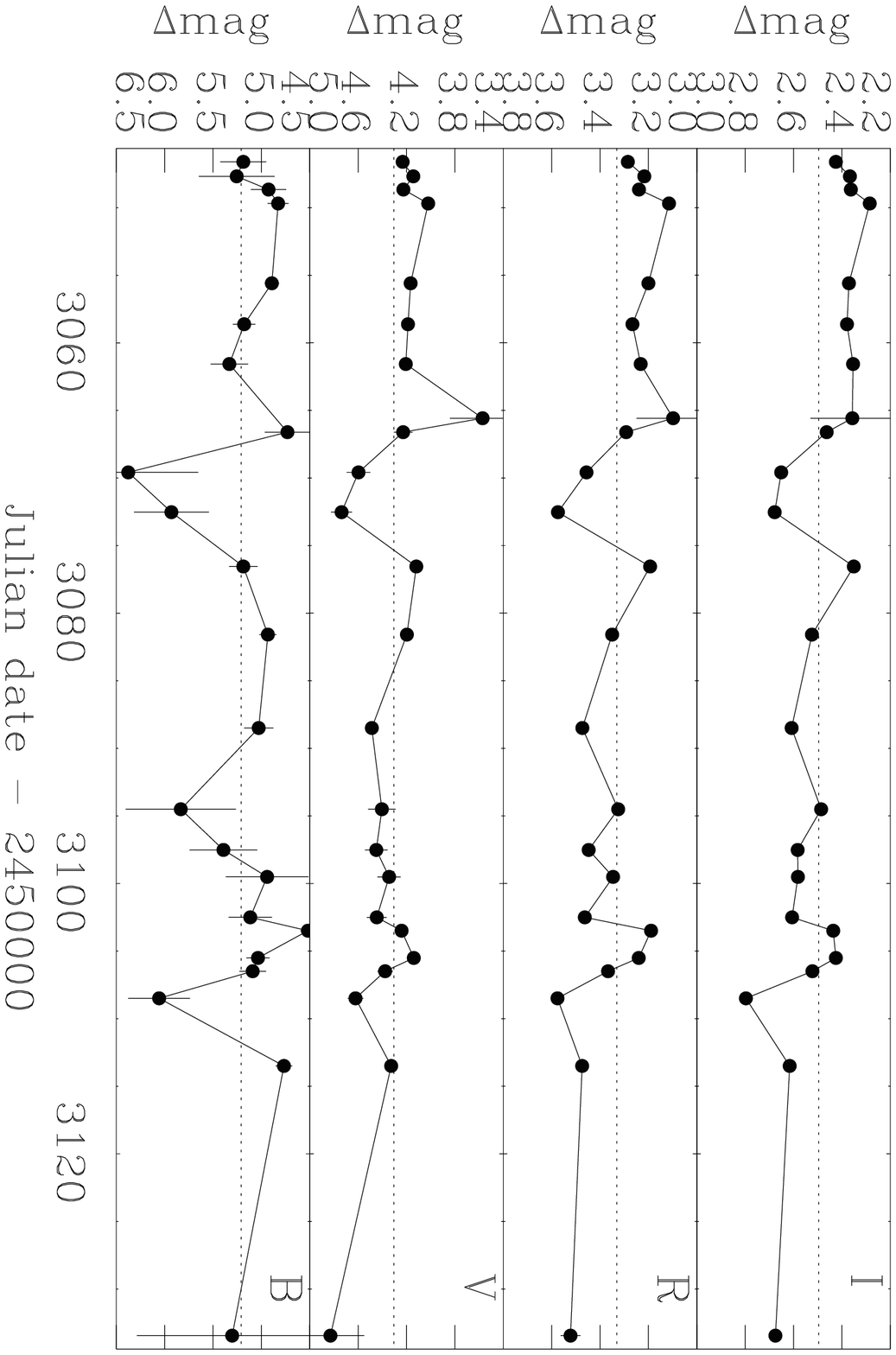}
\end{figure}
 
\begin{figure}
\plotone{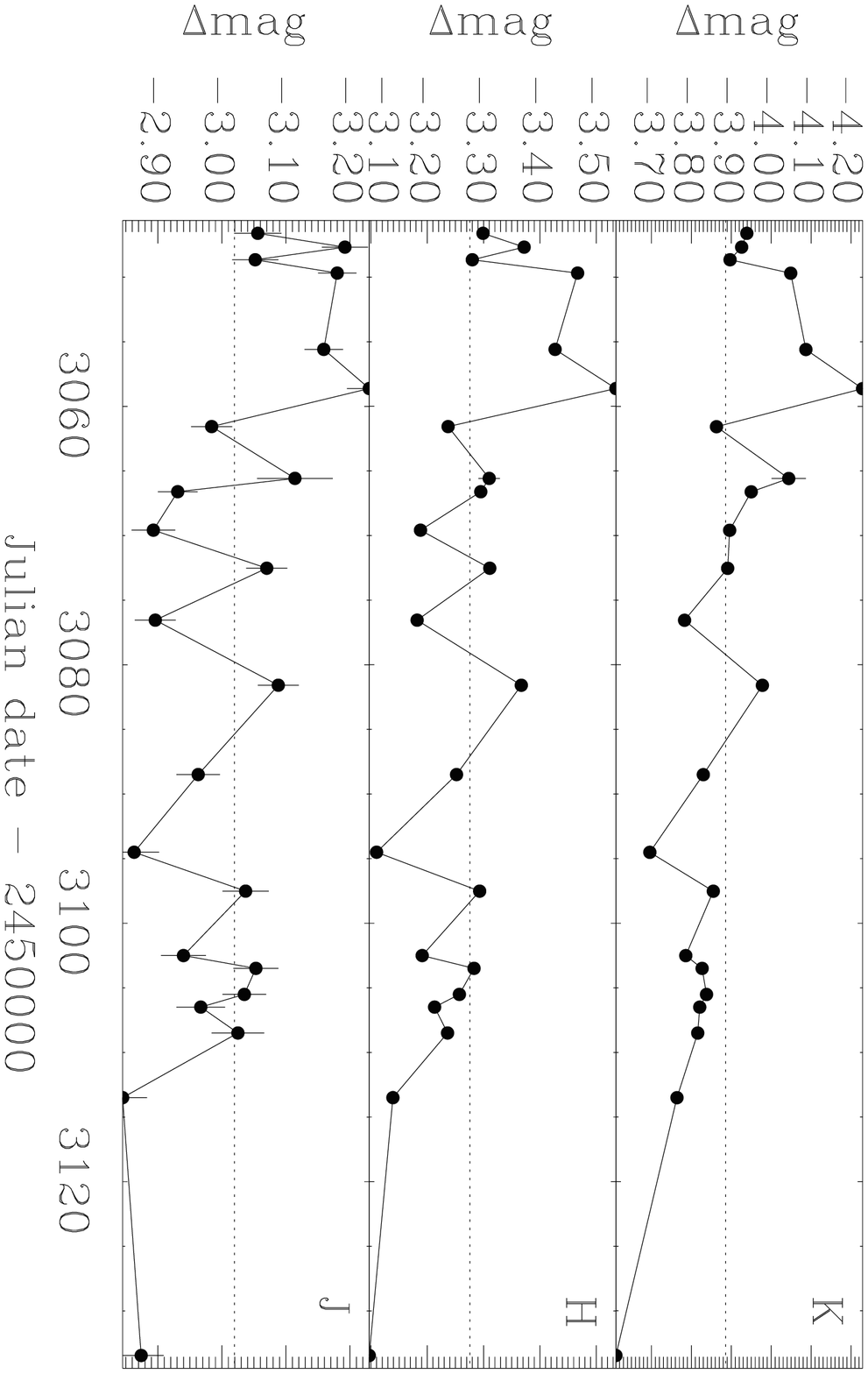}
\end{figure}

\begin{figure}
\plotone{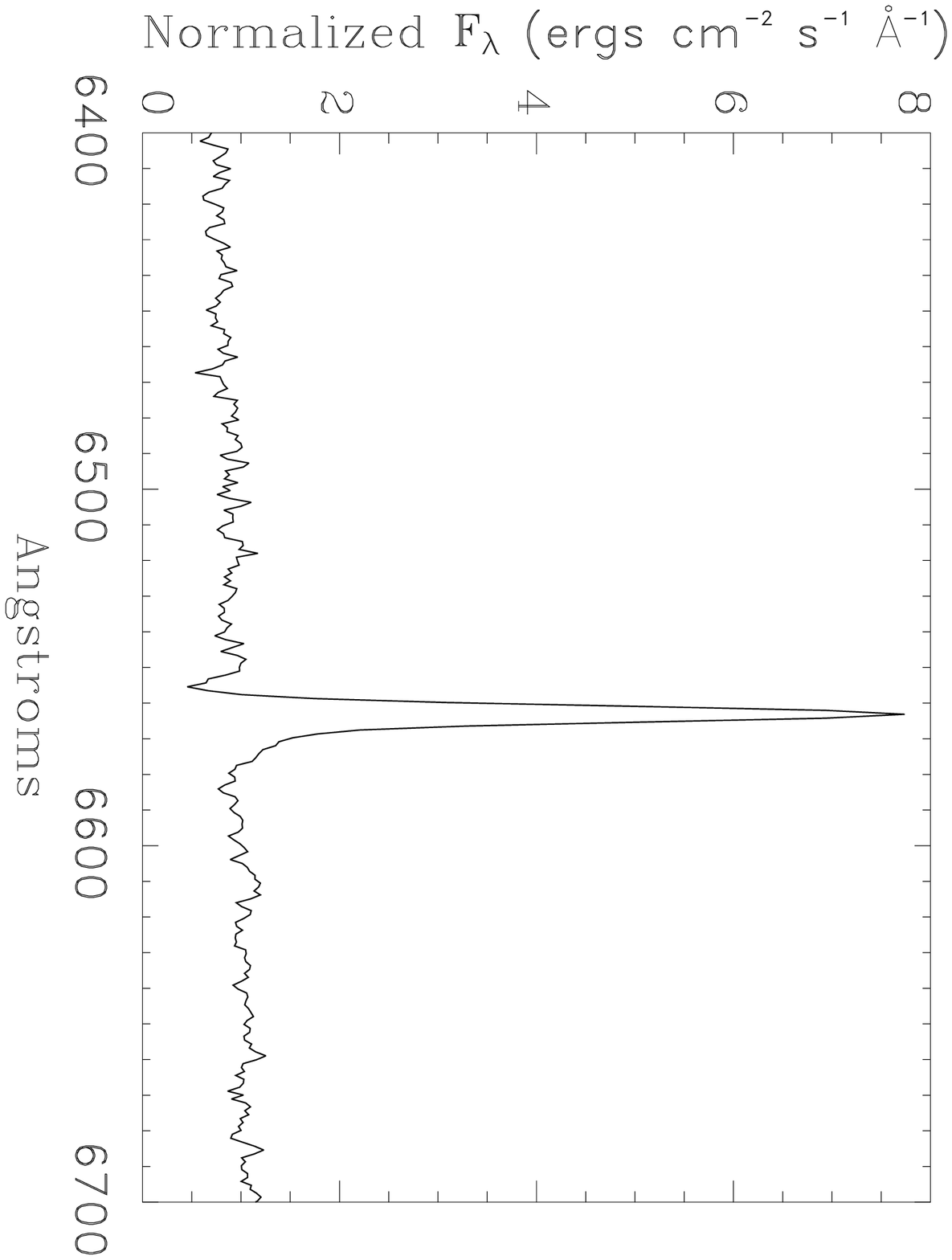}
\end{figure}

\begin{figure}
\plotone{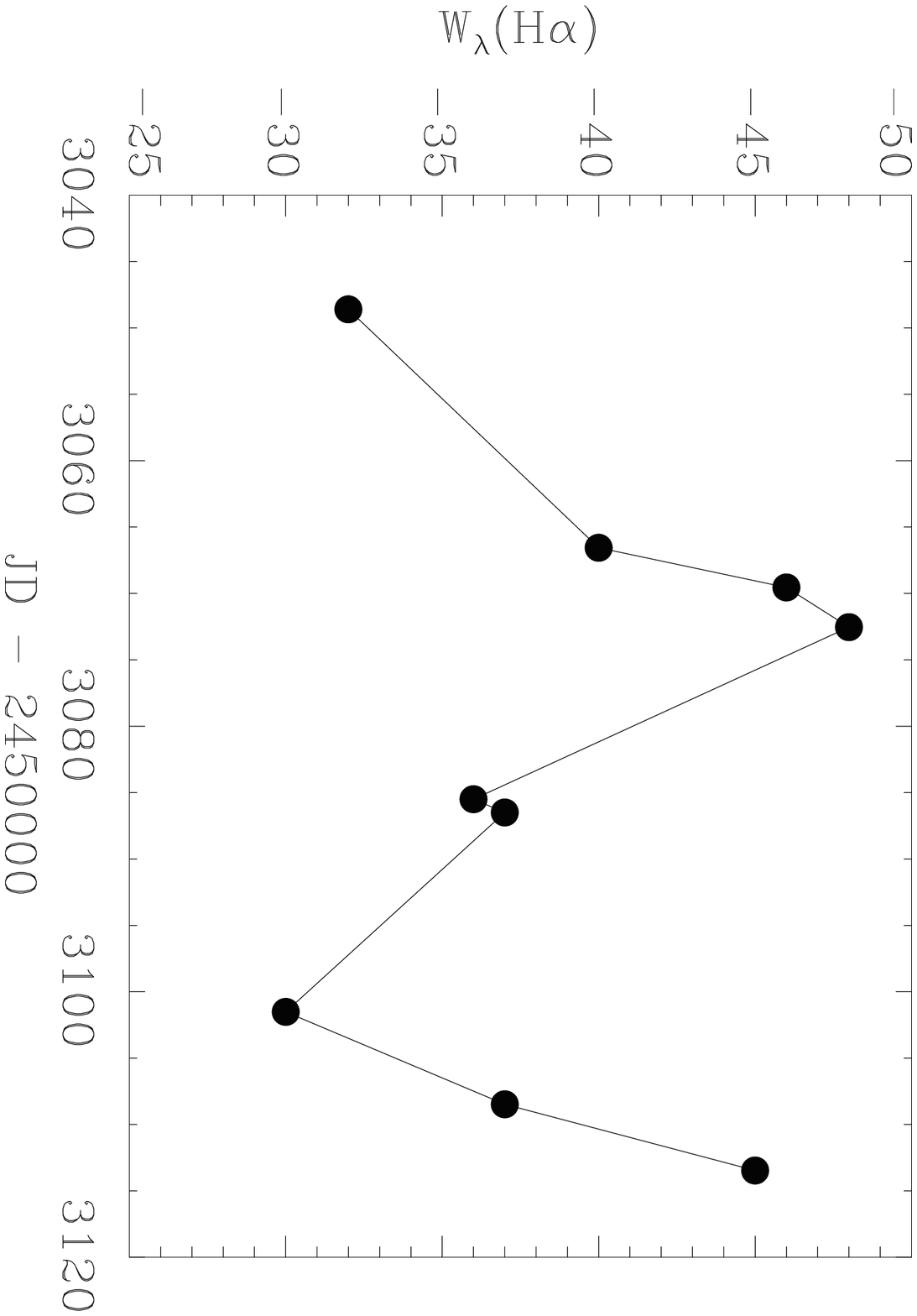}
\end{figure}

\begin{figure}
\plotone{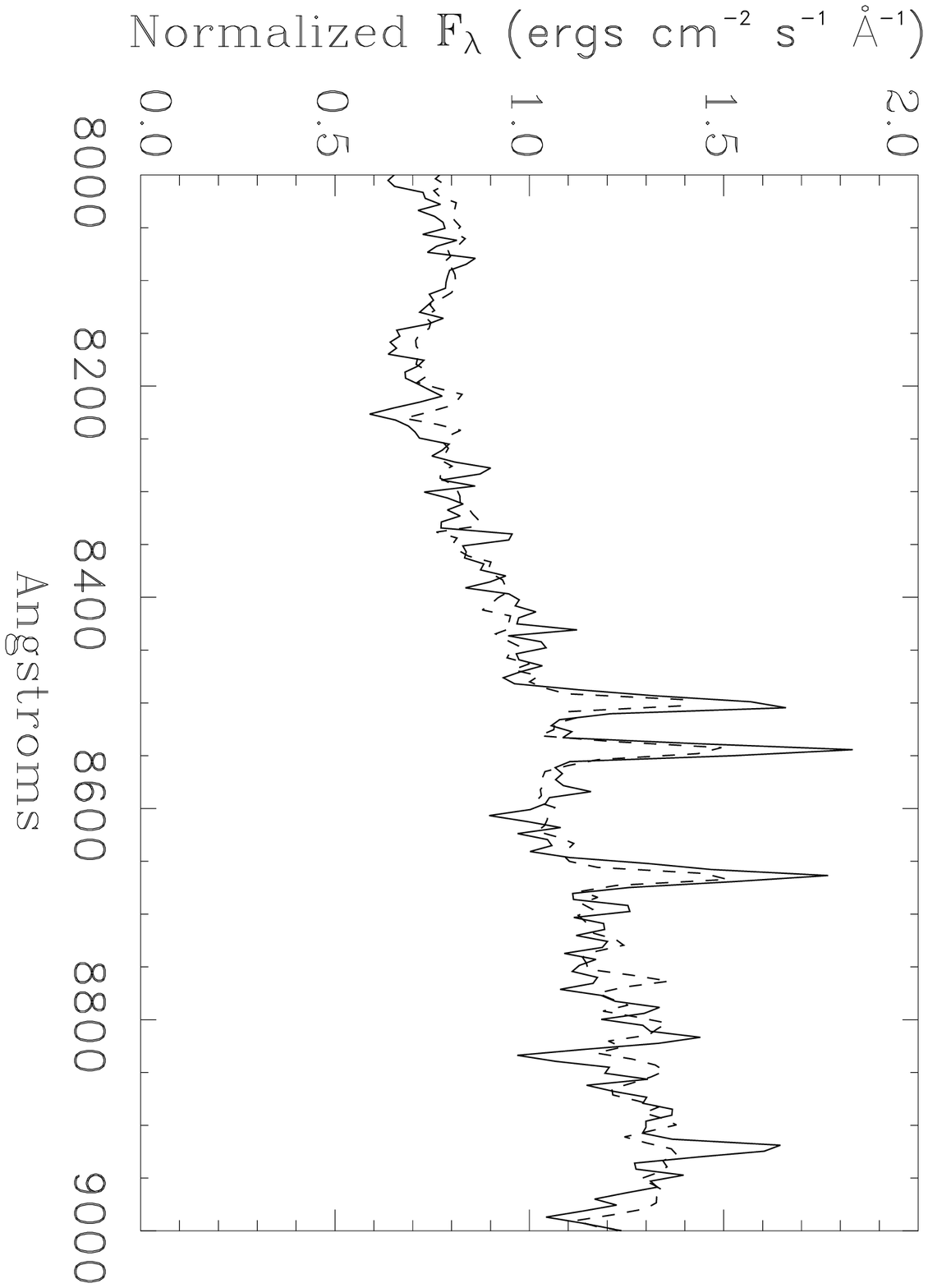}
\end{figure}

\begin{figure}
\plotone{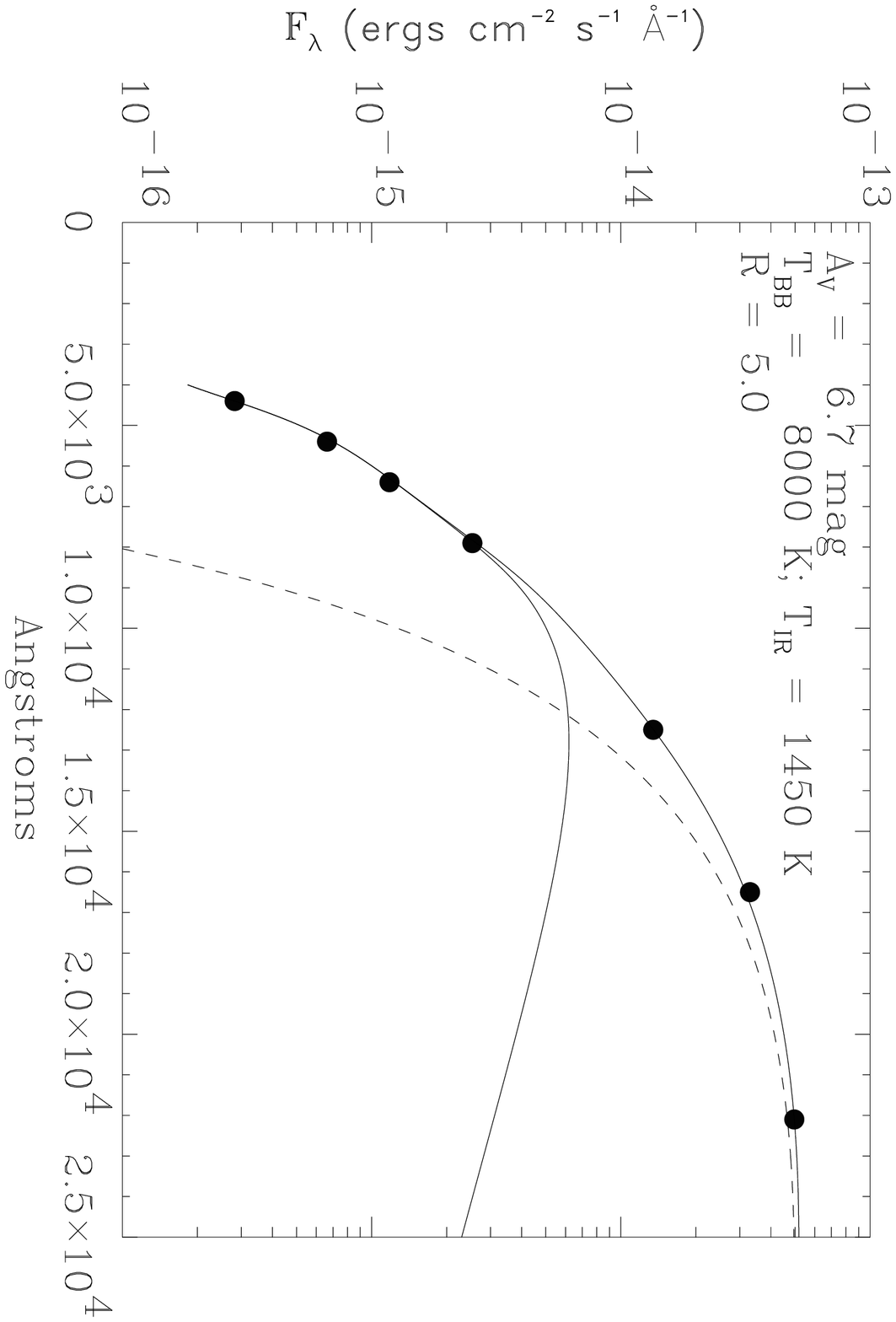}
\end{figure}

\end{document}